\documentclass[10pt]{article}

\usepackage{latexsym}
\usepackage{amsthm}
\usepackage{amsfonts}
\usepackage{amssymb}
\usepackage{amsmath}

\title{On the Minimum Spanning Tree for Directed Graphs with Potential Weights}
\author{V. A. Buslov\thanks{abvabv@bk.ru}\,,
 V. A. Khudobakhshov\thanks{vitaly.khudobakhshov@gmail.com}
 \\ \\ \textit{Department of
Computational Physics} \\ \textit{St.-Petersburg State
University}}

\begin{document}

\maketitle

\begin{abstract}
In general the problem of finding a miminum spanning tree for a weighted
directed graph is difficult but solvable. There are a lot of differences
between problems for directed and undirected graphs, therefore the algorithms
for undirected graphs cannot usually be applied to the directed case. In this
paper we examine the kind of weights such that the problems are equivalent
and a minimum spanning tree of a directed graph may be found by a simple
algorithm for an undirected graph.
\end{abstract}

\section{Introduction}
The problem of finding a minimum spanning tree is well known to graph theorists
as well as to programmers who deal with graph theory applications. For
an undirected case there are a lot of simple algorithms such as Prim's algorithm
\cite{Prim:57}. For directed graphs, a general solution also exists, see papers
\cite{Bock:71, Chu-Liu:65, Edmonds:67} for more information. Furthermore, there
are some optimizations for directed and undirected cases based on Fibonacci
heaps \cite{Gab-Gal-Spe:86}.

In certain problems in physics we deal with directed graphs whose weights
by definition satisfy the given relation. Sometimes the properties of weights
provide a possibility of simplifying the problem to an undirected case. For
example, consider a directed graph whose weights satisfy the equation 
\begin{equation}\label{WeightsEquation}
	Q_{ij} = \varphi_{ij} - \varphi_{ii}.
\end{equation}
The main result of this paper is that for the graph whose a weight matrix is $Q$
with positive entries a minimum spanning tree may be found by a simple algorithm
for a corresponding undirected graph whose a weight matrix is symmetric
matrix $\varphi$.

One can think about the weight $\varphi_{ij}$ as a height of a potential barrier
which has to be surmounted in order to get to point $i$ from point $j$. From
this point of view $\varphi_{ii}$ is a local minimum of a potential well.
So weights which satisfy Equation (\ref{WeightsEquation}) will be called
\emph{potential}.

\section{Definitions}
Depending on the problem, directed trees may be defined in two ways. 
Usually it is supposed that in-degree $id(v) \leq 1$ for all vertices $v$.
However we define tree such that out-degree $od(v) \leq 1$ and $od(v) = 1$
iff $v$ is the root. We will denote $G(V,E,\omega)$ for an undirected graph
where $V$ is a vertex set, $E$ is an edge set and $\omega$ is a weight matrix
corresponding to the edge set. Clearly $\omega$ is always symmetric. In the case
of a directed graph we will use the same notation except a prime to denote
a directed graph and we will use $A$ instead of $E$ in order to note that in
the directed case we will have an arc set. Also in the undirected case the
weight matrix is not neccessary symmetric.   

\section{Minimum Spanning Trees}
Suppose we have an undirected graph $G(V,E,\varphi)$, where $\varphi$ is
a symmetric matrix whose diagonal entries are not neccessary equal to zero,
but inequations $\varphi_{ii} < \varphi_{ik}$ and $\varphi_{ii} < \varphi_{ki}$
hold for all $k \neq i$.  For every $G$ we can define a directed graph
$G'(V,A,Q)$ where $Q$ is a weight matrix which satisfies Equation
(\ref{WeightsEquation}). We must find a tree $T$ which minimizes the following 
expression
\begin{equation}\label{TreeWeight}
	w(T) = \sum_{(i,j) \in T} Q_{ij}.
\end{equation}
This tree is called a minimum spanning tree for the weighted directed graph
$G'$. The minimum spanning tree for undirected graph $G$ is defined in a similar
way.

For this class of weighted directed graphs the following proposition can be
posed.
\newtheorem{teo}{Proposition}
\begin{teo}
A minimum spanning tree of a directed graph $G'$ coincides with a minimum
spanning tree for an undirected graph $G$ with a root in vertex $k$ for which
$\varphi_{kk}$ is minimum.
\begin{proof}
The minumum for (\ref{TreeWeight}) can be written, according for $Q$ weights
properties in following form:
\[
\min_{T \subset G'} w(T) =
\min_{T \subset G'} \sum_{(i,j) \in T} (\varphi_{ij} - \varphi_{ii}).
\]
Allowing for the facts that the number of arcs in spanning tree $T$
equals $|V|-1$ and the off-diagonal entries do not depend on 
diagonal ones, the minimum of the previous expression equals 
\[
\min_{T \subset G' } \sum_{(i,j) \in T} \varphi_{ij} - 
\max_{k} \sum_{i \neq k} \varphi_{ii}.
\]
It is clear that if $\varphi_{ii}=0$ for all $i$ then
$G'$ changes\footnote{
In this case we can change the ordered pair $(i,j)$ to a 2-elements
subset $\{i,j\}$.} to $G$.
Therefore the first summand in the previous expression equals to
the weight of the corresponding a minumum spanning tree of the undirected graph
$G$.
\[
\min_{T \subset G } \sum_{\{i,j\} \in T} \varphi_{ij} - 
\max_{k} \sum_{i \neq k} \varphi_{ii}
\]
It follows
\[
\min_{T \subset G } \sum_{\{i,j\} \in T} \varphi_{ij} - 
\sum_{i = 1}^{|V|} \varphi_{ii} + \min_{k} \varphi_{kk}
\]
As a result, the minimum spanning tree of $G'$ can be given from
the minimum spanning tree of $G$ by fixing a root in vertex $k$ with a minimum
value of $\varphi_{kk}$.
\end{proof}
\end{teo}

\section{Conclusion and Future Work}
In the previous section it is shown that algorithms for undirected graphs
can be applied to a directed case if weights of a directed graph have
certain special properties. In the future we plan to get a solution for a
more difficult problem, finding the minimum spanning forest for graphs
whose weights are potential.

\bibliographystyle{plain} 
\bibliography{mst-preprint}

\end{document}